\documentclass[pra,twocolumn,showpacs,superscriptaddress,amsfonts,amsmath,floatfix]{revtex4}

\usepackage{graphicx}
\usepackage{color}

\begin{document}

\title{Ground state properties of a one-dimensional strongly-interacting
Bose-Fermi mixture in a double-well potential}

\author{K. Lelas}
\affiliation{Faculty of Electrical Engineering Mechanical
Engineering and Naval Architecture, University of Split, Rudjera
Bo\v skovi\' ca BB, 21000 Split, Croatia}
\author{D.~Juki\'{c}}
\affiliation{Department of Physics,
University of Zagreb, Bijeni\v cka c. 32, 10000 Zagreb, Croatia}
\author{H.~Buljan}\email{hbuljan@phy.hr} 
\affiliation{Department of Physics,
University of Zagreb, Bijeni\v cka c. 32, 10000 Zagreb, Croatia}

\date{\today}

\begin{abstract}
We calculate the reduced single-particle density matrix (RSPDM), momentum
distributions, natural orbitals and their occupancies, for a strongly interacting
one-dimensional Bose-Fermi mixture in a double-well potential
with a large central barrier.
For mesoscopic systems, we find that the ground state properties
qualitatively differ for mixtures with even number of particles
(both odd-odd and even-even mixtures) in comparison to mixtures with
odd particle numbers (odd-even and even-odd mixtures).
For even mixtures the momentum distribution is smooth, whereas the momentum
distribution of odd mixtures possesses distinct modulations; the differences
are observed also in the off-diagonal correlations of the RSPDM,
and in the occupancies of natural orbitals.
The calculation is based on a derived formula which enables efficient
calculation of the RSPDM for mesoscopic mixtures in various potentials.
\end{abstract}

\pacs{05.30.-d, 03.75.Ss, 67.85.Pq} \maketitle

\section{Introduction}
\label{sec:intro}

The experiments with interacting ultracold atomic gases loaded in various
external potentials offer great opportunities for probing versatile many-body
states (e.g. see \cite{Bloch2008} for a review), from weakly interacting
gases up to strongly correlated states. Interactions between
the atoms can be tuned in some cases (e.g., by employing Feshbach resonances),
while external potentials can assume various shapes including optical lattices,
elongated and transversely tight traps and many others. Systems of interacting atoms
in double-well potentials exhibit particularly interesting phenomena which were 
studied over the years, for example, a bosonic Josephson junction 
\cite{Milburn1997,Smerzi1997,Albiez2005,Levy2007}, squeezing and entanglement 
of matter waves \cite{Jo2007,Esteve2008}, matter wave interference  
\cite{Andrews1997,Schumm2005}, and recently exact many-body quantum dynamics 
in a one dimensional (1D) quantum well \cite{Sakmann2009}. In this work 
we focus on a strongly interacting Bose-Fermi mixture of particles in a 
1D double well potential.

The achievement of quantum degeneracy in Bose-Fermi mixtures 
\cite{Truscot2001,Schreck2001,Modugno2002,Hadzibabic2002,
Goldwin2004,Silber2005,Ospelkaus2006,Shin2008} has stimulated many studies of 
these systems. 
In 1D geometry, several theoretical approaches explored such mixtures. For 
example, the mean-field approximation has been utilized 
to study phase separation \cite{Das2003}. However, for strongly correlated 
systems, which are more likely to occur in one- than in three-dimensional 
systems, the mean field approach is not appropriate. These systems can 
be studied by using exactly solvable models and/or sophisticated numerical 
calculations. Luttinger liquid theory has been used to study pairing 
instabilities and phase diagrams \cite{Cazalilla2003,Mathey2004}. Numerical calculations 
were used to obtain phase diagram for mixture with unequal masses \cite{Rizzi2008}.
One-dimensional Bose-Fermi mixture with a finite coupling strength and 
without an external trapping potential were studied in Refs. 
\cite{Imambekov2006,Frahm2005,Batchelor2005}.
Recently, an exactly solvable model describing 1D Bose-Fermi mixture with
strong interactions has been studied in Ref. \cite{Girardeau2007}.
The ground state wave functions for arbitrary external potentials were constructed;
in the model, strong ("impenetrable core") interactions are present between 
bosons, bosons and fermions, whereas fermions are mutually noninteracting and 
they are spin-polarized.
The correlation functions including the one-body density matrix were addressed 
for the ring geometry and the harmonic confinement \cite{Girardeau2007}.
The ground state properties and expansion dynamics were further explored 
in Ref. \cite{Fang2009}. 

The solution of the model presented in Ref. \cite{Girardeau2007} follows 
the Fermi-Bose mapping idea to calculate exact wave functions in the 
so called Tonks-Girardeau (TG) model of "impenetrable-core" bosons 
\cite{Girardeau1960}. This model has been experimentally realized several 
years ago \cite{TG2004,Kinoshita2006}, with atoms in tight transversely 
confined atomic waveguides \cite{Olshanii}, at low temperatures, and with 
strong effective interactions \cite{Olshanii,Petrov,Dunjko}. 
Besides the wave functions \cite{Girardeau1960}, the correlation functions
such as the reduced single particle density matrix (RSPDM)
and related quantities including distributions of momenta, natural orbitals
and their occupancies, have been studied for the TG system over the years
\cite{Lenard1964,Korepin1993,Girardeau2001,Papenbrock2003,Forrester2003,Rigol2004,
Rigol2005,Minguzzi2005,Pezer2007,Lelas2007,Goold2008} for the ground states
on the circle \cite{Lenard1964,Forrester2003}, in harmonic confinement
\cite{Girardeau2001,Papenbrock2003,Forrester2003,Rigol2004}, for excited "dark-soliton"
eigenstates \cite{Lelas2007}, in a split-trap potential \cite{Goold2008},
and also for time-dependent states (e.g., see \cite{Rigol2005,Minguzzi2005,Pezer2007}).

Here we study ground state properties of a strongly interacting 
Bose-Fermi mixture in a double-well potential; we use the model from Ref. \cite{Girardeau2007}.  
As a first step, we derive a formula which enables efficient
calculation of the RSPDM for mesoscopic mixtures in various potentials;
it reduces to a calculation of RSPDM for TG bosons \cite{Pezer2007} in an
incoherent mixed state. The formula is employed to calculate the RSPDM, momentum
distributions, natural orbitals and their occupancies, for ground state
of the mixture in a double-well.
For mesoscopic systems, we find that the ground state properties
qualitatively differ for mixtures with even number of particles
(both odd-odd and even-even mixtures) in comparison to mixtures with
odd particle numbers (odd-even and even-odd mixtures).
For even mixtures the momentum distribution is smooth, whereas the momentum
distribution of odd mixtures possesses modulations; the differences
are observed also in the off-diagonal correlations of the RSPDM,
and in the occupancies of natural orbitals.

\section{The model}
\label{sec:mod}

We study a mixture of $N_B$ bosons and $N_F$ spin-polarized fermions in 
one-dimensional geometry, in a ground state of an external
potential $V(x)$. Bosons and fermions experience the same
external potential, and their masses are assumed to be
approximately equal.
This condition can be satisfied for combination of bosonic and fermionic
isotopes of the same element, such as $^{39(41)}K-^{40}K$, or $Yb$ with 
several stable isotopes, and $^{86(84)}Rb-^{87(85)}Rb$ (for detailed discussion, 
see \cite{Imambekov2006} and references therein).
Bosons interact via a very strong repulsive contact potential, that is, 
their interaction is in the Tonks-Girardeau regime. Fermions are mutually 
noninteracting. Interaction between bosons and fermions is also a very strong
repulsive contact potential. The total number of particles is denoted with
$N=N_B+N_F$. The ground state for this system for finite but very strong
interactions can be approximately written as \cite{Girardeau2007}:
\begin{align}
&\psi(x_{F1},\ldots,x_{FN_F},x_{B1},\ldots,x_{BN_B})= \nonumber \\
&\prod_{1\leq i < j \leq N_B} \mbox{sgn}(x_{Bj}-x_{Bi})
\prod_{j=1}^{N_B} \prod_{i=1}^{N_F}\mbox{sgn}(x_{Bj}-x_{Fi}) \nonumber \\
&\psi_S(x_{F1},\ldots,x_{N_F},x_{B1},\ldots,x_{N_B}). 
\label{GS}
\end{align}
Here,
\begin{equation}
\psi_{S}(x_{1},\ldots,x_{N})=\frac{1}{\sqrt{N!}} \sum_{P\in
S_N}(-)^P \phi_{Pj_1}(x_1)\ldots \phi_{Pj_N}(x_N),
\label{Slater}
\end{equation}
denotes a Slater determinant wave function constructed from the
single particle wave functions $\phi_j(x)$, which are the $N$
lowest single-particle eigenstates of the potential $V(x)$:
\begin{equation}
-\frac{d^2 \phi_j(x)}{dx^2}+V(x)\phi_j(x)=E_j \phi_j(x),
\end{equation}
$j=1,2,\ldots,N=N_F+N_B$. In Eq. (\ref{Slater}), $P$ denotes a 
permutation from the group $S_N$. 
For the clarity of the exposition, it is convenient to define the
following quantities. Let us consider a subset of single particle
states, chosen from the set $\{\phi_j(x)|j=1,\ldots,N \}$ by
crossing out $k$ single-particle states; let
$J=\{j_1,j_2,\ldots,j_k \}$ denote the indices of the crossed
states, and $L=\{l_1,l_2,\ldots,l_{N-k} \}$ the indices of the remaining
states (obviously $J\cap L=\emptyset$, $J\cup L=\{1,\ldots,N\}$).
We define the Slater determinant state
\begin{align}
&\psi_{S}^{(j_1,j_2,\ldots,j_k)}(x_{1},\ldots,x_{N-k})=\frac{1}{\sqrt{(N-k)!}} \nonumber \\
& \times \sum_{P\in S_{N-k}}(-)^P \phi_{Pl_1}(x_1)\ldots \phi_{Pl_{N-k}}(x_{N-k}),
\end{align}
where $P$ is a permutation of indices $(l_1,l_2,\ldots,l_{N-k})$.
Thus, the indices upon $\psi_{S}^{(j_1,j_2,\ldots,j_k)}$ denote
the crossed out states, rather than the ones used in the Slater
determinant.

Let $\psi_{TG}^{(j_1,j_2,\ldots,j_k)}=A\psi_{S}^{(j_1,j_2,\ldots,j_k)}$
denote a symmetric Tonks-Girardeau state obtained by acting with a
unit antisymmetric function $A=\prod_{1\leq i<j\leq N-k}
\mbox{sgn}(x_j-x_i)$ upon $\psi_{S}^{(j_1,j_2,\ldots,j_k)}$. The
RSPDM of the state $\psi_{TG}^{(j_1,j_2,\ldots,j_k)}$ will be
denoted by $\rho_{TG}^{(j_1,j_2,\ldots,j_k)}$,
\begin{align}
& \rho_{TG}^{(j_1,j_2,\ldots,j_k)}(x,y) =  (N-k) \int dx_{2} \ldots dx_{N-k} \nonumber \\
& \; \; \; \times \psi_{TG}^{(j_1,j_2,\ldots,j_k)}(x,x_{2}\ldots,x_{N-k})^*  \nonumber \\
& \; \; \; \; \; \; \; \; \; \; \times 
\psi_{TG}^{(j_1,j_2,\ldots,j_k)}(y,x_{2}\ldots,x_{N-k}).
\end{align}
The quantities $\psi_S$, $\psi_{TG}$, and $\rho_{TG}$ etc. will
refer to states and the correlation functions obtained from the
full set of single particle states $\{\phi_j(x)|j=1,\ldots,N \}$.

\section{Formula for the RSPDM}
\label{sec:rspdm}

We are interested in properties of the ground state of the
Bose-Fermi mixture described by the state (\ref{GS}). To
explore the one-particle observables of this Bose-Fermi mixture, we need
to construct the RSPDM of the bosonic and the fermionic
subsystems, respectively. The one-body density matrix for
the bosonic part of the mixture is defined as
\begin{align}
& \eta_{N_B,N_F}(x,y) =  N_B \int 
dx_{F1} \ldots dx_{FN_F}
dx_{B2} \ldots dx_{BN_B} \nonumber\\
& \; \; \; \; \; \; \; \; \times 
  \psi^*(x_{F1},\ldots,x_{FN_F},x,x_{B2},\ldots,x_{BN_B})\nonumber\\
& \; \; \; \; \; \; \; \; \; \; \; \;  \times 
  \psi(x_{F1},\ldots,x_{FN_F},y,x_{B2},\ldots,x_{BN_B}).
\end{align}
It is straightforward to verify that $\eta_{N_B,N_F}(x,y)$ can be
calculated from the RSPDM describing $N$ Tonks-Girardeau bosons,
$\eta_{N_B,N_F}(x,y)=N_B/N \rho_{TG}(x,y)$. This correlation
function can be efficiently calculated by using the procedure
derived in Ref. \cite{Pezer2007}.

The calculation of the density matrix describing the fermionic
component of the mixture,
\begin{align}
& \mu_{N_B,N_F}(x,y) =  N_F \int 
dx_{F2} \ldots dx_{FN_F}
dx_{B1} \ldots dx_{BN_B} \nonumber\\
& \; \; \; \; \; \; \; \; \times 
  \psi^*(x,x_{F2},\ldots,x_{FN_F},x_{B1},\ldots,x_{BN_B})\nonumber\\
& \; \; \; \; \; \; \; \; \; \; \; \;  \times 
  \psi(y,x_{F2},\ldots,x_{FN_F},x_{B1},\ldots,x_{BN_B}),
\label{fermiRSPDM}
\end{align}
is much more involved. The derivation can be reduced to
calculations of the RSPDM for an incoherent mixed TG state as follows:
\begin{widetext}
\begin{align}
\mu_{N_B,N_F}(x,y)  
 & = \frac{N_F}{N!} \int dx_{F2} \dots dx_{FN_F}
                        dx_{B1} \dots dx_{BN_B} \nonumber \\
& \times \prod_{i=1}^{N_B} \mbox{sgn}(x-x_{Bi})\mbox{sgn}(y-x_{Bi}) 
   \left| \begin{array}{ccc}
        \phi_1^*(x)& \ldots & \phi_N^*(x) \\
        \phi_1^*(x_{F2})& \ldots & \phi_N^*(x_{F2}) \\
        \vdots  & \ddots & \vdots \\
        \phi_1^*(x_{FN_F}) & \ldots & \phi_N^*(x_{FN_F}) \\
        \phi_1^*(x_{B1}) & \ldots & \phi_N^*(x_{B1}) \\
        \vdots  & \ddots & \vdots \\
        \phi_1^*(x_{BN_B}) & \ldots & \phi_N^*(x_{BN_B}) \\
    \end{array} \right|
    \left| \begin{array}{ccc}
        \phi_1(y)& \ldots & \phi_N(y) \\
        \phi_1(x_{F2})& \ldots & \phi_N(x_{F2}) \\
        \vdots  & \ddots & \vdots \\
        \phi_1(x_{FN_F}) & \ldots & \phi_N(x_{FN_F}) \\
        \phi_1(x_{B1}) & \ldots & \phi_N(x_{B1}) \\
        \vdots  & \ddots & \vdots \\
        \phi_1(x_{BN_B}) & \ldots & \phi_N(x_{BN_B}) \\
    \end{array} \right|.
\label{rspdm_F}
\end{align}
\end{widetext}
The determinants above can be expanded along their 2nd row
according to the Laplace formula, after which the integral over
$x_{F2}$ is trivially performed:
\begin{widetext}
\begin{align}
\mu_{N_B,N_F}(x,y)  & = \frac{N_F}{N!} \int 
     dx_{F3} \dots dx_{FN_F}
     dx_{B1} \dots dx_{BN_B} \nonumber \\ \times 
  & \prod_{i=1}^{N_B} \mbox{sgn}(x-x_{Bi})\mbox{sgn}(y-x_{Bi}) 
  \sum_{j,l=1}^N (-)^{2+j} (-)^{2+l} 
  \int  \phi_j^*(x_{F2}) \phi_l(x_{F2}) dx_{F2} \nonumber \\ \times 
   &\left| \begin{array}{cccccc}
        \phi_1^*(x) & \cdots & \phi_{j-1}^*(x)  & \phi_{j+1}^*(x)  &  \cdots & \phi_N^*(x) \\
        \phi_1^*(x_{F3}) & \cdots & \phi_{j-1}^*(x_{F3})  & \phi_{j+1}^*(x_{F3})  &  \cdots & \phi_N^*(x_{F3}) \\
        \vdots  &  & \vdots & \vdots & & \vdots \\
        \phi_1^*(x_{FN_F}) & \cdots & \phi_{j-1}^*(x_{FN_F})  & \phi_{j+1}^*(x_{FN_F})  &  \cdots & \phi_N^*(x_{FN_F}) \\
        \phi_1^*(x_{B1}) & \cdots & \phi_{j-1}^*(x_{B1})  & \phi_{j+1}^*(x_{B1})  &  \cdots & \phi_N^*(x_{B1}) \\
        \vdots  &  & \vdots & \vdots & & \vdots \\
        \phi_1^*(x_{BN_B}) & \cdots & \phi_{j-1}^*(x_{BN_B})  & \phi_{j+1}^*(x_{BN_B})  &  \cdots & \phi_N^*(x_{BN_B}) \\
    \end{array} \right| \nonumber \\ \times
   &\left| \begin{array}{cccccc}
        \phi_1(y)& \cdots & \phi_{l-1}(y)  & \phi_{l+1}(y)  &  \cdots & \phi_N(y) \\
        \phi_1(x_{F3})& \cdots & \phi_{l-1}(x_{F3})  & \phi_{l+1}(x_{F3})  &  \cdots & \phi_N(x_{F3}) \\
        \vdots  &  & \vdots & \vdots & & \vdots \\
        \phi_1(x_{FN_F}) & \cdots & \phi_{l-1}(x_{FN_F})  & \phi_{l+1}(x_{FN_F})  &  \cdots & \phi_N(x_{FN_F}) \\
        \phi_1(x_{B1}) & \cdots & \phi_{l-1}(x_{B1})  & \phi_{l+1}(x_{B1})  &  \cdots & \phi_N(x_{B1}) \\
        \vdots  &  & \vdots & \vdots & & \vdots \\
        \phi_1(x_{BN_B}) & \cdots & \phi_{l-1}(x_{BN_B})  & \phi_{l+1}(x_{BN_B})  &  \cdots & \phi_N(x_{BN_B}) \\
    \end{array} \right|
=\frac{N_F}{N(N_F-1)} \sum_{j=1}^N \mu_{N_B,N_F-1}^{(j)}(x,y),
\end{align}
\end{widetext}
where we have used $\int\phi_j^*(x_{F2}) \phi_l(x_{F2}) dx_{F2}=\delta_{jl}$.
The index $j$ in $\mu_{N_B,N_F-1}^{(j)}$ means that the
single-particle state $\phi_j$ has been crossed out from the
Slater determinant used in the formula for the ground state, see
Sec. \ref{sec:mod}. Thus, we have derived a recursion formula
which reduces the calculation of the fermionic RSPDM
$\mu_{N_B,N_F}(x,y)$, to the calculation of $N$ fermionic correlation functions,
$\mu_{N_B,N_F-1}(x,y)$ (with one fermion less in the mixture). By successively
applying the recursive formula above, it is straightforward to
obtain
\begin{align}
& \mu_{N_B,N_F}(x,y) = \frac{N_F! (N_B+1)!}{N!} \nonumber \\ 
& \; \; \; \; \; \; \; \; \; \; \; \; \times \sum_{1\leq j_1<\ldots<j_{N_F-1}\leq N}
\mu_{N_B,1}^{(j_1,\ldots,j_{N_F-1})}(x,y) \nonumber \\
& = \frac{N_F! N_B!}{N!} \sum_{1\leq j_1<\ldots<j_{N_F-1}\leq N}
\rho_{TG}^{(j_1,\ldots,j_{N_F-1})}(x,y). \label{final}
\end{align}
The correlation functions $\mu_{N_B,1}^{(j_1,\ldots,j_{N_F-1})}(x,y)$ correspond to 
a mixture with one fermion and $N_B$ bosons; it is straightforward to see from the
definition of $\mu_{N_B,N_F}$ that $\mu_{N_B,1}^{(j_1,\ldots,j_{N_F-1})} \propto
\rho_{TG}^{(j_1,\ldots,j_{N_F-1})}$, i.e., the system with just
one extra fermion has identical RSPDM (up to a normalization constant) to the
system with $N_B+1$ TG bosons placed in the proper single particle
orbitals.

Thus, the density matrix $\mu_{N_B,N_F}(x,y)$ is equal (up to a proportionality 
constant) to a sum of density matrices of TG states from an ensemble.  
Each TG state from the ensemble describe $N_B+1$ TG bosons; these states are 
constructed by choosing $N_B+1$ orbitals from the full set 
of single particle states $\{\phi_j(x)|j=1,\ldots,N \}$. Apparently, there 
are ${N \choose N_B+1}$ such states, i.e., there are ${N \choose N_B+1}$ terms 
in the sum (\ref{final}). 
Thus, the density matrix $\mu_{N_B,N_F}(x,y)$ is equivalent to the density matrix of
$N_B+1$ bosons in a mixed TG state; the mixed state is an
incoherent superposition of the ground state and many excited TG states, 
each of which is constructed by some choice of $N_B+1$ orbitals as stated above. 
The calculation thus reduces to applying the algorithm of Ref.
\cite{Pezer2007}. Numerical calculation becomes too time consuming 
if the number of terms in the sum (\ref{final}) is too large. 
Nevertheless, it can be performed efficiently for mesoscopic
systems.

From the RSPDM, one can extract observables such as the momentum distribution,
and important quantities like the natural orbitals (NOs) and their
occupancies. For example, the fermionic momentum distribution is
given by
\begin{equation}
n_{\mu}(k)=\frac{1}{2 \pi} \int dx dy e^{ik(x-y)} \mu_{N_B,N_F}(x,y);
\label{nFK}
\end{equation}
the eigenfunctions of the fermionic RSPDM, $\Phi_{\mu,i}(x)$, are called the natural orbitals (NOs),
\begin{equation}
\int dx \mu_{N_B,N_F}(x,y) \Phi_{\mu,i}(x) = \lambda_{\mu,i} \Phi_{\mu,i}(y), \;\;\; i=1,2,\ldots;
\end{equation}
the eigenvalues $\lambda_{\mu,i}$ are the occupancies of these orbitals.
The bosonic momentum distribution $n_{\eta}(k)$, NOs $\Phi_{\eta,i}(x)$, and
occupancies $\lambda_{\eta,i}$, are defined by using equivalent relations for the
bosonic RSPDM.

\section{Mixture in a split trap}
\label{sec:trap}

In this section we apply the presented formalism to study the RSPDM, momentum
distribution, natural orbitals and their occupancies for a Bose-Fermi mixture in
a double well potential of the form
\begin{eqnarray}
V(x) & = & V_{ho}(x)+V_G(x) \nonumber \\
 & = & \nu^2 x^2+V_{0} e^{-(x/\sigma)^2}
 \label{potencijal}
\end{eqnarray}
where the parameter $\sigma$ ($V_{0}$) denotes the width (height, respectively) of the
Gaussian barrier which splits the harmonic potential. We note in
passing that the presented result may depend on the shape of the
double-well potential (e.g., for the split-box potential) as
will be discussed below. Here we work in dimensionless units; the
connection to physical units can be made straightforwardly: For
example, if $m$ is the mass of bosons (which is equal to the mass 
of fermions), and $x_{unit}$ the unit of space (which can be chosen at will), 
then the unit of energy is $E_{unit}=\hbar^2/(2mx_{unit}^2)$.

\begin{figure}
\centerline{
\mbox{\includegraphics[width=0.45\textwidth]{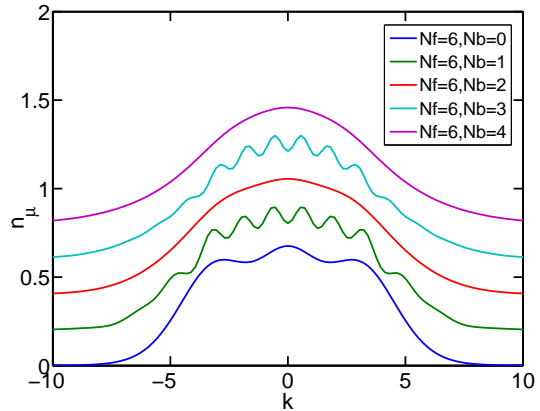}}
}
\caption{
(color online) Momentum distributions $n_{\mu}(k)$ for $N_F=6$ (all curves)
and $N_B=0,1,2,3,4$; curves are ordered from bottom to top and shifted by a
constant for better visibility.}
\label{MD-Nf6Nb0-4}
\end{figure}

\begin{figure}
\centerline{
\mbox{\includegraphics[width=0.45\textwidth]{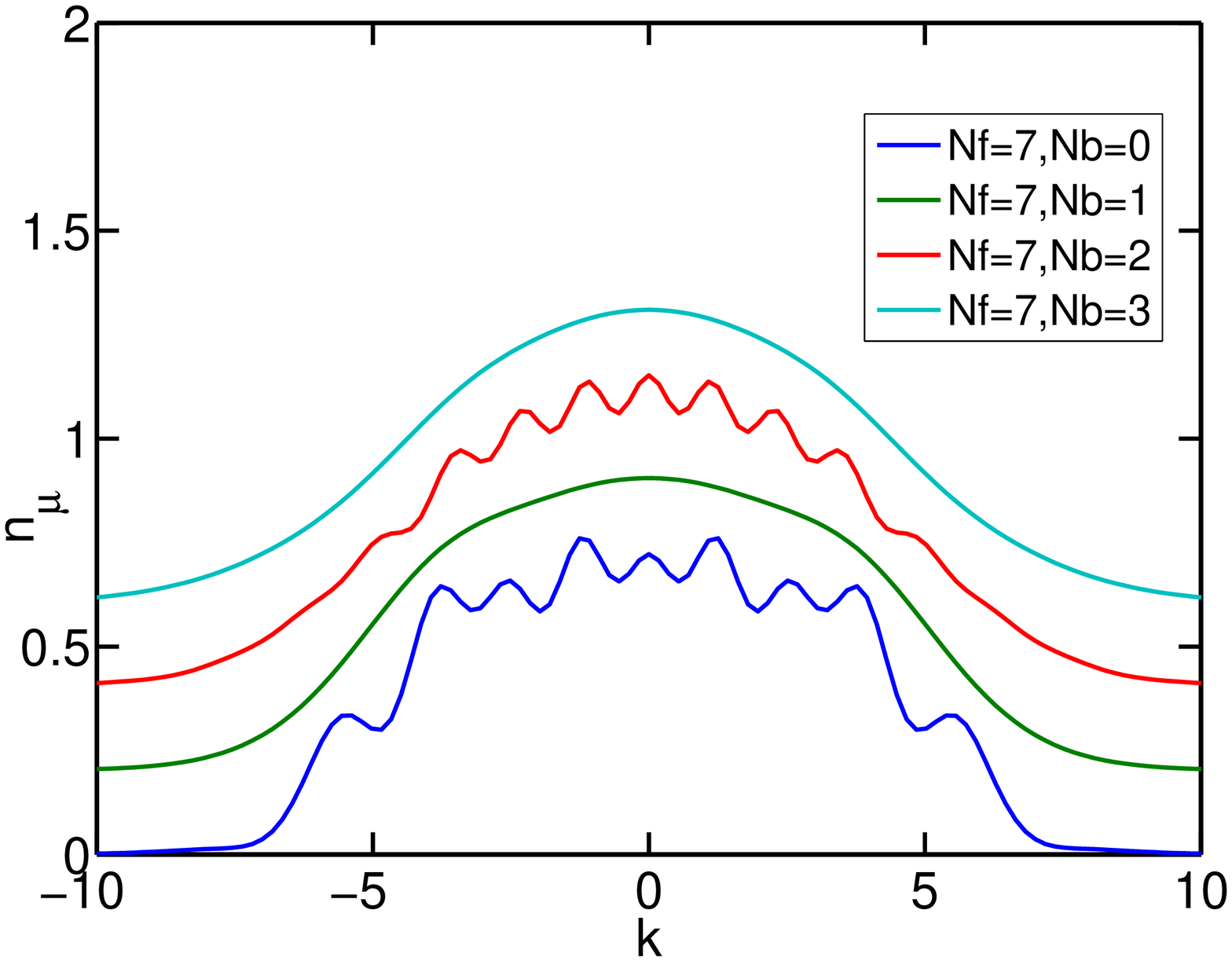}}
}
\caption{
(color online) Momentum distributions $n_{\mu}(k)$ for $N_F=7$ (all curves)
and $N_B=0,1,2,3$: curves are ordered from bottom to top and shifted by a
constant for better visibility.}
\label{MD-Nf7Nb0-3}
\end{figure}

\begin{figure}
\centerline{
\mbox{\includegraphics[width=0.45\textwidth]{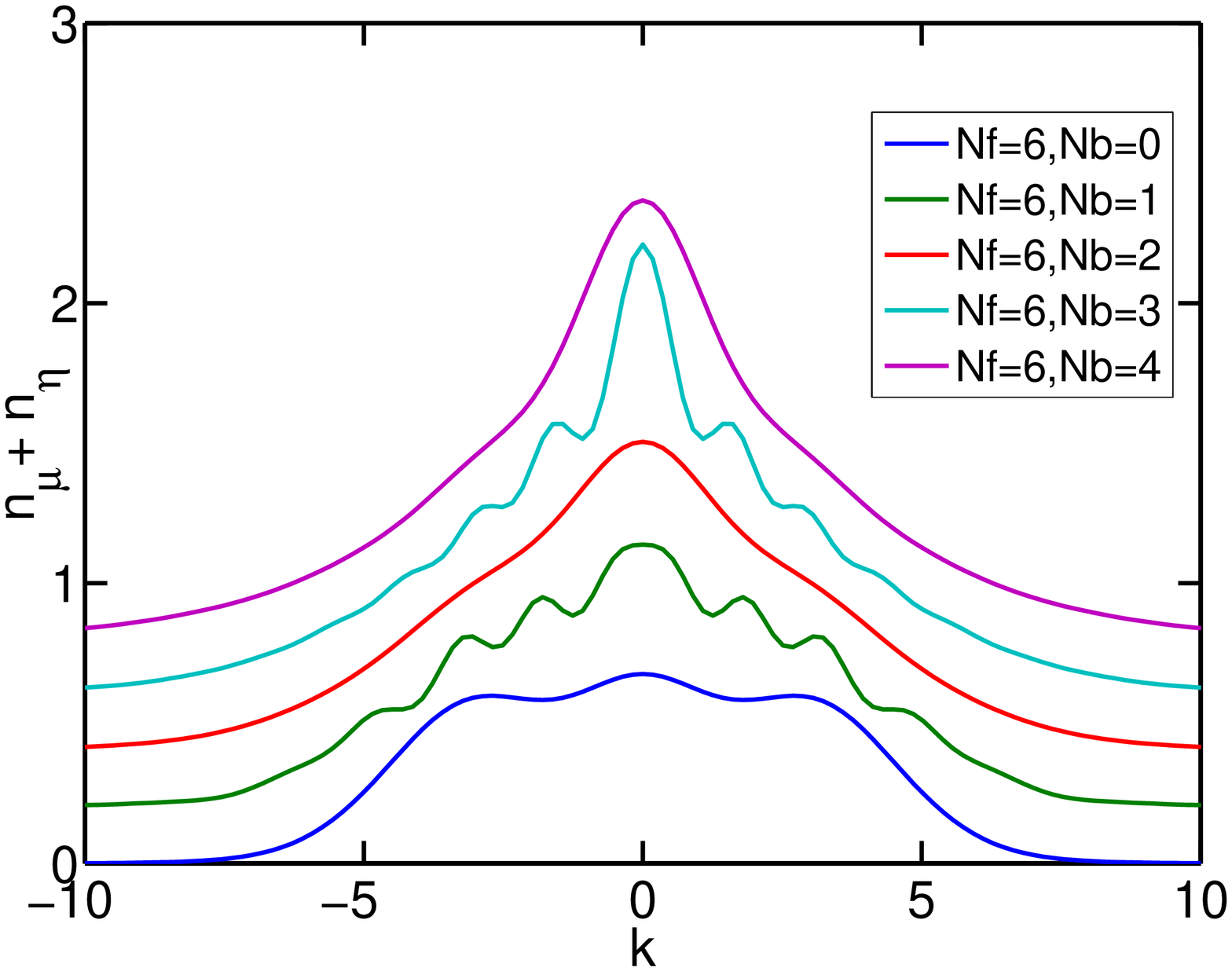}}
}
\caption{
(color online) The total momentum distributions $n_{\mu}(k)+n_{\eta}(k)$ for
$N_F=6$ (all curves) and $N_B=0,1,2,3,4$; curves are ordered from bottom to top
and shifted by a constant for better visibility.
}
\label{MDTotNf6}
\end{figure}

\begin{figure}
\centerline{
\mbox{\includegraphics[width=0.45\textwidth]{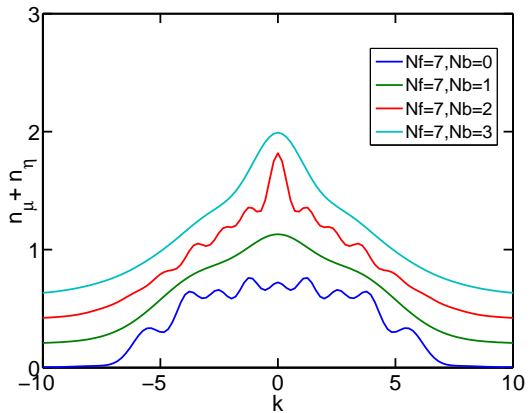}}
}
\caption{
(color online) The total momentum distributions $n_{\mu}(k)+n_{\eta}(k)$ for
$N_F=7$ (all curves) and $N_B=0,1,2,3$; curves are ordered from bottom to top
and shifted by a constant amount for better visibility.
}
\label{MDTotNf7}
\end{figure}

In what follows we present result of numerical simulations for the
double-well parameters $\sigma=0.3$ and $V_{0}=230$; the harmonic
trap frequency parameter is $\nu^2=7.5$. It should be emphasized
that we have focused our attention to the cases when the splitting
potential is sufficiently high, say, when $V_{0}$ is at least
several times larger than the energy of the $N$th single-particle
state of the potential $V_{ho}(x)$. Thus, all our conclusions should be
understood to hold in this limit. The momentum distribution of the
bosonic component was shown to be equivalent to that of $N$ TG
bosons placed within $V(x)$ and this case has already been
addressed in Ref. \cite{Goold2008}. Thus, we first turn our
attention to the behavior of the fermionic momentum distribution
in the mixture $n_{\mu}(k)$, in dependence of the number of
particles. Figure \ref{MD-Nf6Nb0-4} shows $n_{\mu}(k)$ for $N_F=6$
and $N_B=0,1,2,3,4$, while Fig. \ref{MD-Nf7Nb0-3} shows the same
quantity for $N_F=7$ and $N_B=0,1,2,3$. From these figures we
observe a qualitative difference in behavior of the fermionic
momentum distribution in dependence on the parity of the total
number of particles $N=N_F+N_B$: If $N$ is even, then $n_{\mu}(k)$
has a smooth bell-shaped profile. In contrast, when $N$ is odd,
then there are nonnegligible modulations on top of the bell-shaped
profile of $n_{\mu}(k)$; in our simulations, for the parameters
presented here, we find that the number of peaks in $n_{\mu}(k)$
for odd $N$ is the same as the number of
fermions in the mixture $N_F$ plus two additional humps on the
side bands of the distribution. This parity dependent behavior is
reflected onto the behavior of the total momentum distributions
[$n_{\mu}(k)+n_{\eta}(k)$] which are displayed in Figures
\ref{MDTotNf6} and \ref{MDTotNf7} for the same combinations of
particles as presented in Figs. \ref{MD-Nf6Nb0-4} and
\ref{MD-Nf7Nb0-3}. Interestingly, odd-odd combinations yield
results very similar to the even-even ones, whereas even-odd
combination yields results similar to the odd-even combinations.
Thus, the parity of the {\em total} number of particles determines
the behavior of the fermionic component, at least for the
mesoscopic numbers of particles studied here.

\begin{figure}
\centerline{
\mbox{\includegraphics[width=0.45\textwidth]{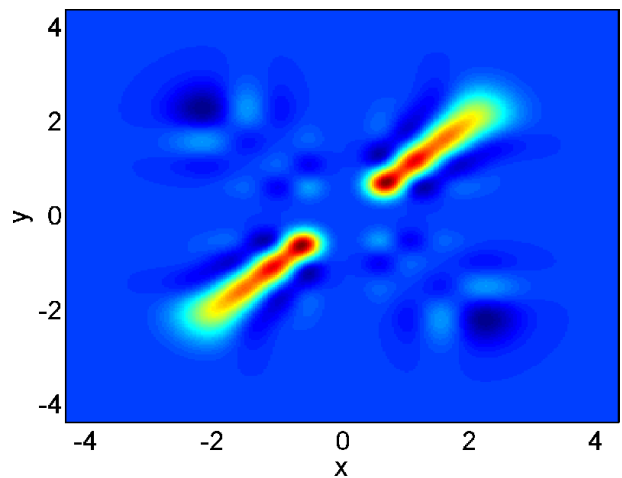}}
}
\caption{
(color online) Contour plot of the fermionic RSPDM for $N_F=6$ and $N_B=1$.
}
\label{RSPDM61}
\end{figure}

\begin{figure}
\centerline{
\mbox{\includegraphics[width=0.45\textwidth]{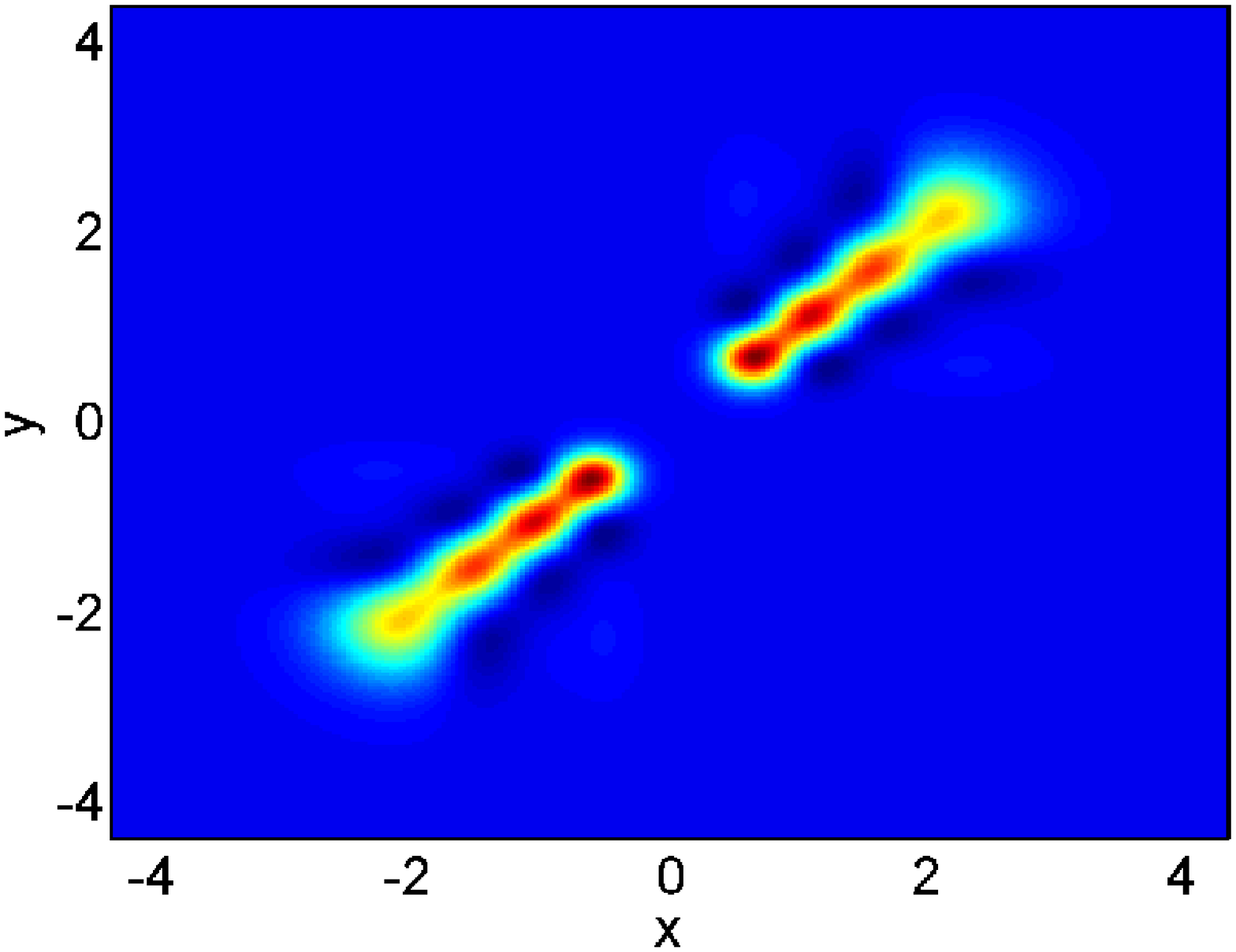}}
}
\caption{
(color online) Contour plot of the fermionic RSPDM for $N_F=6$ and $N_B=2$.
}
\label{RSPDM62}
\end{figure}

\begin{figure}
\centerline{
\mbox{\includegraphics[width=0.45\textwidth]{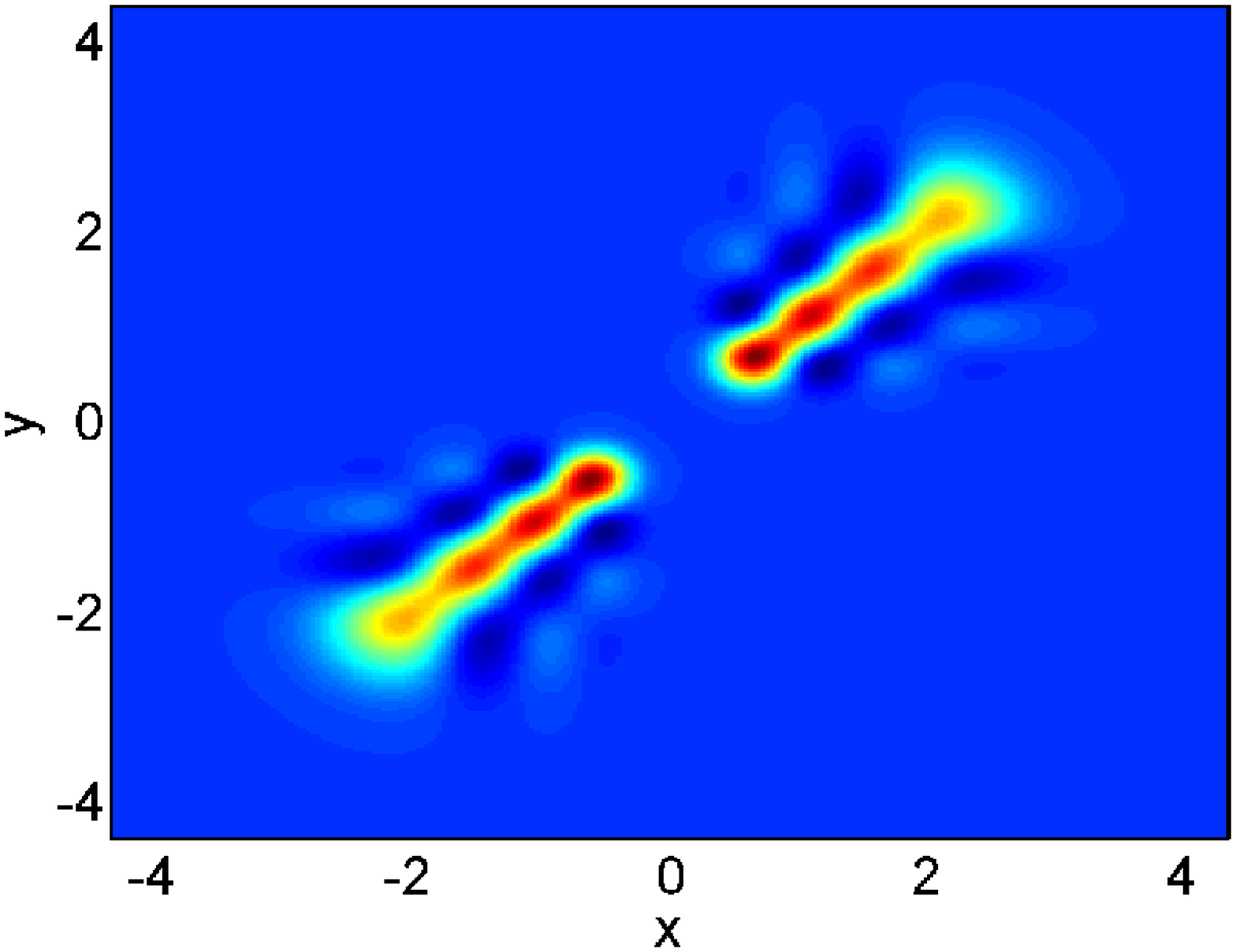}}
}
\caption{
(color online) Contour plot of the fermionic RSPDM for $N_F=7$ and $N_B=1$.
}
\label{RSPDM71}
\end{figure}

\begin{figure}
\centerline{
\mbox{\includegraphics[width=0.45\textwidth]{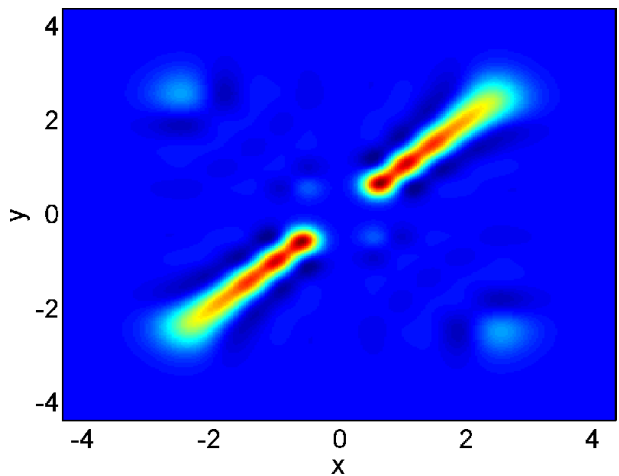}}
}
\caption{
(color online) Contour plot of the fermionic RSPDM for $N_F=7$ and $N_B=2$.
}
\label{RSPDM72}
\end{figure}

The behavior of the fermionic momentum distribution results from the
properties of the fermionic density matrix $\mu_{N_B,N_F}(x,y)$ which
is illustrated in Fig. \ref{RSPDM61} for $N_F=6$ and $N_B=1$,
Fig. \ref{RSPDM62} for $N_F=6$ and $N_B=2$,
Fig. \ref{RSPDM71} for $N_F=7$ and $N_B=1$,
and in Fig. \ref{RSPDM72} for $N_F=7$ and $N_B=2$.
From these figures we observe that the most significant difference between
the total even and odd numbers of particles occurs in the second and the 
fourth quadrant of the $x$-$y$ plane: If $N$ is even, the values of the
fermionic RSPDM for $x<0<y$ and $y<0<x$ are negligible, $\mu_{N_B,N_F}(x,y)\approx 0$;
In contrast to that, if $N$ is odd, there are some oscillations of RSPDM in 
the second and the fourth quadrant, in particular close to the line 
$\mu_{N_B,N_F}(x,-x)$. These observations indicate that there is much greater
spatial coherence between the fields at the two sides of the well for odd $N$.
A similar observation has been made for the TG gas in a split trap \cite{Goold2008}.

\begin{figure}
\centerline{
\mbox{\includegraphics[width=0.45\textwidth]{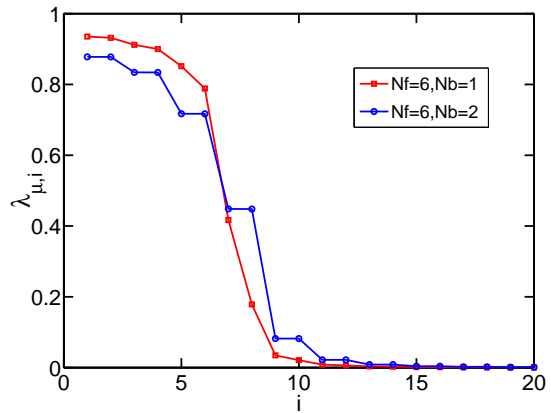}}
}
\caption{
(color online) The occupancies of the fermionic RSPDM for the combinations
$N_F=6$ and $N_B=1$ (squares), and $N_F=6$ and $N_B=2$ (circles).
The lines serve to guide the eye. 
}
\label{lambdas612}
\end{figure}

\begin{figure}
\centerline{
\mbox{\includegraphics[width=0.45\textwidth]{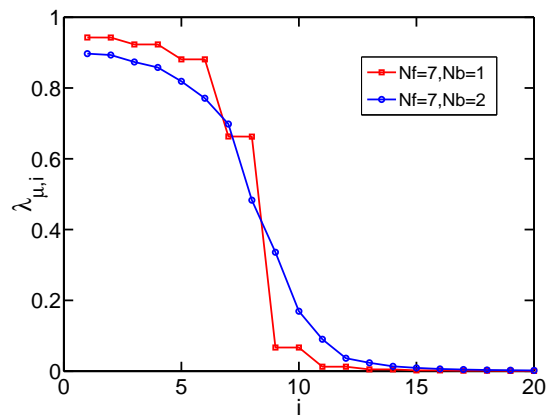}}
}
\caption{
(color online) The occupancies of the fermionic RSPDM for the combinations
$N_F=7$ and $N_B=1$ (squares), and $N_F=7$ and $N_B=2$ (circles).
The lines serve to guide the eye. 
}
\label{lambdas712}
\end{figure}

Let us now focus on the natural orbitals, that is, their occupancies.
The occupancies $\lambda_{\mu,i}$ corresponding to the four fermionic
density matrices from Figs. \ref{RSPDM61}-\ref{RSPDM72}
are illustrated in Figs. \ref{lambdas612} and \ref{lambdas712}.
We immediately observe that when the total number of particles is even,
the occupancies come in pairs and they correspond to the
degenerate natural orbitals. Namely for the total even numbers
of particles the symmetry of the system with respect to the
double well is naturally preserved. However, for the total odd number of
particles this is not the case, which is reflected in the
occupancies which do not come in degenerate pairs but rather decrease
one by one.

The observation on the off diagonal behavior of the fermionic RSPDM 
can be underpinned analytically. To this end, let us assume that 
the trap is split by an infinitely strong delta function, i.e., the 
the mixture is in the potential $V(x)=V_{\mbox{trap}}+\kappa \delta(x)$, where 
$\kappa\rightarrow \infty$; the potential $V_{\mbox{trap}}$ can be 
a harmonic oscillator trap, or it may have some other functional form. 
For the strongly-interacting Bose-Fermi mixture in a ground state of such a 
potential we can prove the following: 
If the total number of particles $N$ is even, then $\mu_{N_B,N_F}(x,y)=0$ 
in the second and the fourth quadrant of the $x$-$y$ plane, that is, for 
$x<0<y$ and $x>0>y$. 

Consecutive single-particle states in a split-trap potential are degenerate, 
i.e., $E_{2m-1}=E_{2m}$ for $m=1,2,\ldots$; moreover, degenerate eigenstates 
are simply related: $\phi_{2m}(x)=\mbox{sgn}(x) \phi_{2m-1}(x)$. By using this 
relation, one of the two Slater determinants (\ref{Slater}) which enter the formula 
for the RSPDM (\ref{fermiRSPDM}) [the determinant which depends on variable $x$] 
can be written as 
\begin{widetext}
\begin{align}
\psi_S= \left( \begin{array}{ccccc}
\phi_{1}(x)     & \mbox{sgn}(x)\phi_{1}(x)      & \cdots & \phi_{N-1}(x)   & \mbox{sgn}(x)\phi_{N-1}(x)     \\
\phi_{1}(x_2)   & \mbox{sgn}(x_2)\phi_{1}(x_2)  & \cdots & \phi_{N-1}(x_2) & \mbox{sgn}(x_2)\phi_{N-1}(x_2) \\
 \vdots         & \vdots                        &        & \vdots        & \vdots                       \\
\phi_{1}(x_N)   & \mbox{sgn}(x_N)\phi_{1}(x_N)  & \cdots & \phi_{N-1}(x_N) & \mbox{sgn}(x_N)\phi_{N-1}(x_N)
\end{array} \right).
\label{det1}
\end{align}
\end{widetext}
Here we have simplified the notation and labeled the coordinates 
as $x_j$, where $j$ runs up to the total number of particles $N$, 
that is, we do not explicitly refer to the fermionic 
or bosonic coordinates, as it is redundant for the proof. 
Let us assume that $x<0<y$, and that 
$N_1$ coordinates are negative, $N_2$ are positive, $N_1>N_2$, and 
$N_1+N_2=N$. First, let us demonstrate that the determinant (\ref{det1}) 
is zero in this case. 
In order to see that, we assume that $x,x_2,\ldots,x_{N_1}<0$, and that  
the rest of the coordinates are positive (due to the antisymmetry of 
the Slater determinant, any other choice of positive and negative coordinates 
would yield the same result). 
Let us add the first column to the second one in (\ref{det1}), then 
the third to the fourth column and so on to obtain 
\begin{widetext}
\begin{align}
\psi_S= \left( \begin{array}{ccccc}
\phi_{1}(x)     & 0      & \cdots & \phi_{N-1}(x)   & 0    \\
\phi_{1}(x_2)   & 0      & \cdots & \phi_{N-1}(x_2) & 0    \\
 \vdots         & \vdots       &        & \vdots        & \vdots                      \\
\phi_{1}(x_{N_1})     & 0      & \cdots & \phi_{N-1}(x_{N_1})   & 0       \\
\phi_{1}(x_{N_1+1})   & 2\phi_{1}(x_{N_1+1})  & \cdots & \phi_{N-1}(x_{N_1+1}) & 2\phi_{N}(x_{N_1+1}) \\
 \vdots         & \vdots       &        & \vdots        & \vdots                       \\
\phi_{1}(x_N)   & 2\phi_{1}(x_N)  & \cdots & \phi_{N-1}(x_N) & 2\phi_{N-1}(x_N)
\end{array} \right).
\label{det2}
\end{align}
\end{widetext}
Thus, the first $N_1$ entries of every even column, $2,4,\ldots,N$, is zero. 
Suppose that we shift all these even columns all the way to the rigth;
the determinant is then proportional to 
\begin{widetext}
\begin{align}
\left( \begin{array}{ccccccc}
\phi_{1}(x)     & \phi_{3}(x)  & \cdots      & \phi_{N-1}(x)   & 0 & \cdots  & 0    \\
\phi_{1}(x_2)   & \phi_{3}(x_2)  & \cdots    & \phi_{N-1}(x_2) & 0 & \cdots  & 0    \\
 \vdots         & \vdots    &       &   \vdots   & \vdots      &   & \vdots         \\
\phi_{1}(x_{N_1})     & \phi_{3}(x_{N_1})  &  \cdots   & \phi_{N-1}(x_{N_1}) & 0  & \cdots & 0   \\
\phi_{1}(x_{N_1+1})   & \phi_{3}(x_{N_1+1})  & \cdots & \phi_{N-1}(x_{N_1+1}) & 2\phi_{1}(x_{N_1+1}) & \cdots & 2\phi_{N-1}(x_{N_1+1}) \\
 \vdots         & \vdots  &     &   \vdots     & \vdots    &    & \vdots             \\
\phi_{1}(x_N)   & \phi_{3}(x_N) &  \cdots & \phi_{N-1}(x_N) & 2\phi_{1}(x_N) & \cdots & 2\phi_{N-1}(x_N)
\end{array} \right).
\label{det3}
\end{align}
\end{widetext}
In the upper right corner there is a block of zeros with the size 
$N_1\times N/2$. Since $N_1>N/2$, the determinant is identically zero. 
In fact, it is straightforward to verify by using the procedure outlined above 
that the determinant is exactly zero whenever $N_1\neq N_2$, as long as 
$x$ and $y$ have opposite signs and $N$ is even. 
On the other hand, if $N_1=N_2$, then the other Slater 
determinant entering the formula for the RSPDM in Eq. (\ref{fermiRSPDM}) 
[the one which depends on $y$ variable], does not have equal number 
of positive and negative coordinates and therefore it is zero, which 
completes our proof. It is straightforward to see that the same result holds 
for the bosonic part of the mixture, and therefore for the mixture
as a whole. This result can be interpreted as follows: 
If two slits were opened on the opposite sides of the barrier, 
and if the gas was allowed to drop from the slits, expand, and interfere, 
the interference fringes would have not been observed for even $N$ 
and sufficiently high barrier. In fact, such an experiment could be used to 
determine the parity of the mixture. 

We note that this results corresponds to the observation 
made in Ref. \cite{Goold2008} on the difference in spatial coherence of the 
Tonks-Girardeau gas in a split-trap in dependence of the parity of the 
number of TG bosons. Our proof covers this case as well. 

Finally, it is worthy to note that the shape of the external potential
can influence the presented results. For example, for a box potential
with a Gaussian as a split-barrier we did not find the differences
between the even and odd numbers of particles presented above, that is, 
in all cases the momentum distribution was smooth.
For a potential of the form $\propto |x|$ with a splitting Gaussian term
the differences between the odd and even numbers are recovered.

\section{Conclusion}
\label{sec:conc}

In conclusion, we have studied a 1D Bose-Fermi mixture, where the
boson-boson and boson-fermion interactions are very strong and repulsive, 
whereas (spin-polarized) fermions are mutually noninteracting;
the atoms in each species have approximately the same mass.
The ground state for this system for finite but very strong interactions
in an external potential has been constructed in Ref. \cite{Girardeau2007}.
We have studied the ground state properties of the mixture in a double-well trap,
with a sufficiently high barrier. 
More specifically, a formula for the calculation of the reduced one-body density matrix
of the fermionic (and also bosonic) component has been derived, which was subsequently
employed to study the momentum distribution, natural orbitals and their occupancies.
We have found that the behavior of the momentum distribution depends on the 
parity of the total number of particles: 
For even mixtures the momentum distribution is smooth, whereas the momentum
distribution of odd mixtures possesses distinct modulations. 
When the total number of particles is even, the correlations expressed by the 
reduced one-body density matrix are negligible between the two wells. 

\acknowledgments

This work is supported by the Croatian Ministry of Science
(Grant No. 119-0000000-1015). H.B. and D.J. acknowledge support 
from the Croatian-Israeli project cooperation and the Croatian National 
Foundation for Science. K.L. would like to thank Miss Antonela 
Ozreti\' c Banovac for support.




\begin{thebibliography}{99}


\bibitem{Bloch2008}
I. Bloch, J. Dalibard, and W. Zwerger,
Rev. Mod. Phys. {\bf 80}, 885 (2008).


\bibitem{Milburn1997}
G.J. Milburn, J. Corney, E.M. Wright, D.F. Walls,
Phys. Rev. A {\bf 55}, 4318 (1997).

\bibitem{Smerzi1997}
A. Smerzi, S. Fantoni, S. Giovanazzi, and S.R. Shenoy,
Phys. Rev. Lett. {\bf 79}, 4950 (1997).

\bibitem{Albiez2005}
M. Albiez,R. Gati, J. F\" olling, S. Hunsmann, M. Cristiani, and M.K. Oberthaler,
Phys. Rev. Lett. {\bf 95}, 010402 (2005).

\bibitem{Levy2007}
S. Levy, E. Lahoud, I. Shomroni, and J. Steinhauer, 
Nature (London) {\bf 449}, 579 (2007).

\bibitem{Jo2007}
G.-B. Jo, Y. Shin, S. Will, T.A. Pasquini, M. Saba, W. Ketterle,
D.E. Pritchard, M. Vengalattore, and M. Prentiss
Phys. Rev. Lett. {\bf 98}, 030407 (2007).

\bibitem{Esteve2008}
J. Esteve, C. Gross, A. Weller, S. Giovanazzi, M.K. Oberthaler
Nature (London) {\bf 455}, 1216 (2008).

\bibitem{Andrews1997}
M.R. Andrews, C.G. Townsend, H.-J. Miesner, D.S. Durfee, D.M. Kurn, and W. Ketterle,
Science {\bf 275}, 637 (1997).

\bibitem{Schumm2005}
T. Schumm, S. Hofferberth, L.M. Andersson, S. Wildermuth,
S. Groth, I. Bar-Joseph, J. Schmiedmayer, and P. Kr\" uger, 
Nature Physics {\bf 1}, 57 (2005).

\bibitem{Sakmann2009}
K. Sakmann, A.I. Streltsov, O.E. Alon, and L.S. Cederbaum,
arXiv:0905.0902v1.


\bibitem{Truscot2001}
A.G. Truscot, K.E. Strecker, W.I. Alexander, G.B. Partridge, and R.G. Hulet,
Science {\bf 291}, 2570 (2001).

\bibitem{Schreck2001}
F. Schreck, L. Khaykovich, K.L. Corwin, G. Ferrari, T. Bourdel, J.
Cubizolles, and C. Salomon, Phys. Rev. Lett. {\bf 87}, 080403
(2001).

\bibitem{Modugno2002}
G. Modugno, G. Roati, F. Riboli, F. Ferlaino, R. J. Brecha, and M. Inguscio,
Phys. Rev. Lett. {\bf 297}, 2240 (2002).

\bibitem{Hadzibabic2002}
Z. Hadzibabic, C.A. Stan, K. Dieckmann, S. Gupta, M.W. Zwierlein,
A. Gorlitz and W. Ketterle,
Phys. Rev. Lett. {\bf 88}, 160401 (2002).

\bibitem{Goldwin2004}
J. Goldwin, S. Inouye, M. L. Olsen, B. Newman,
B. D. DePaola, and D. S. Jin,
Phys. Rev. A {\bf 70}, 021601(R) (2004).

\bibitem{Silber2005}
C. Silber, S. G\"unther, C. Marzok, B. Deh, P.W. Courteille, and C. Zimmermann,
Phys. Rev. Lett. {\bf 95}, 170408 (2005).

\bibitem{Ospelkaus2006}
S. Ospelkaus, C. Ospelkaus, L. Humbert, K. Sengstock, and K. Bongs,
Phys. Rev. Lett. {\bf 97}, 120403 (2006);

\bibitem{Shin2008}
Y. Shin, A. Schirotzek, C.H. Schunk, and W. Ketterle,
Phys. Rev. Lett. {\bf 101}, 070404 (2008).



\bibitem{Das2003}
K.K. Das, Phys. Rev. Lett. {\bf 90}, 170403 (2003).

\bibitem{Cazalilla2003}
M.A. Cazalilla and A.F. Ho,
Phys. Rev. Lett. {\bf 91}, 150403 (2003).

\bibitem{Mathey2004}
L. Mathey, D.-W. Wang, W. Hofstetter, M.D. Lukin, and E. Demler,
Phys. Rev. Lett. {\bf 93}, 120404 (2004).

\bibitem{Rizzi2008}
M. Rizzi and A. Imambekov,
Phys. Rev. A {\bf 77}, 023621 (2008).

\bibitem{Imambekov2006}
A. Imambekov and E. Demler, Ann. Phys. (N.Y.) 321,
2390 (2006).

\bibitem{Frahm2005}
H. Frahm and G. Palacios,
Phys. Rev. A {\bf 72}, 061604(R) (2005)

\bibitem{Batchelor2005}
M.T. Batchelor, M. Bortz, X.W. Guan, and N. Oelkers,
Phys. Rev. A {\bf 72}, 061603(R) (2005).

\bibitem{Girardeau2007}
M.D. Girardeau and A. Minguzzi,
Phys. Rev. Lett. {\bf 99}, 230402 (2007).

\bibitem{Fang2009}
B. Fang, P. Vignolo, C. Miniatura, and A. Minguzzi,
Phys. Rev. A {\bf 79}, 023623 (2009).



\bibitem{Girardeau1960}
M. Girardeau, J. Math. Phys. {\bf 1}, 516 (1960).

\bibitem{TG2004}
T. Kinoshita, T. Wenger, and D.S. Weiss, Science {\bf 305}, 1125
(2004); B. Paredes, A. Widera, V. Murg, O. Mandel, S. F\" olling,
I. Cirac, G. V. Shlyapnikov, T. W. H\" ansch, and I. Bloch, Nature
(London) {\bf 429}, 277 (2004).

\bibitem{Kinoshita2006}
T. Kinoshita, T. Wenger, and D.S. Weiss, Nature (London) {\bf
440}, 900 (2006).

\bibitem{Olshanii}
M. Olshanii, Phys. Rev. Lett. {\bf 81}, 938 (1998).

\bibitem{Petrov}
D.S. Petrov, G.V. Shlyapnikov, and J.T.M. Walraven, 
Phys. Rev. Lett. {\bf 85} 3745 (2000).

\bibitem{Dunjko}
V. Dunjko, V. Lorent, and M. Olshanii, 
Phys. Rev. Lett. {\bf 86} 5413 (2001).

\bibitem{Lenard1964}
A. Lenard, J. Math. Phys. {\bf 5}, 930 (1964);
T.D. Schultz, ibid. {\bf 4}, {\bf 666} (1963).

\bibitem{Korepin1993}
V.E. Korepin, N.M. Bogoliubov, and A.G. Izergin,
\textit{Quantum Inverse Scattering Method and Correlation Functions}
(Cambridge, Cambridge University Press, 1993).

\bibitem{Girardeau2001}
M.D. Girardeau, E.M. Wright, and J.M. Triscari,
Phys. Rev. A {\bf 63}, 033601 (2001);
G. J. Lapeyre, M. D. Girardeau, and E.M. Wright,
ibid. {\bf 66}, 023606 (2002).

\bibitem{Papenbrock2003}
T. Papenbrock, Phys. Rev. A {\bf 67}, 041601(R) (2003).

\bibitem{Forrester2003}
P. J. Forrester, N. E. Frankel, T. M. Garoni, and N. S. Witte,
Phys. Rev. A 67, 043607 (2003).

\bibitem{Rigol2004}
M. Rigol and A. Muramatsu,
Phys. Rev. A {\bf 70}, 031603(R) (2004).

\bibitem{Rigol2005}
M. Rigol and A. Muramatsu,
Phys. Rev. Lett. {\bf 94}, 240403 (2005).

\bibitem{Minguzzi2005}
A. Minguzzi and D.M. Gangardt,
Phys. Rev. Lett. {\bf 94}, 240404 (2005).

\bibitem{Pezer2007}
R. Pezer and H. Buljan, Phys. Rev. Lett. {\bf 98}, 240403 (2007).

\bibitem{Lelas2007}
H. Buljan, K. Lelas, R. Pezer, and M. Jablan,
Phys. Rev. A, {\bf 76}, 043609 (2007)

\bibitem{Goold2008}
J. Goold and Th. Busch,
Phys. Rev. A {\bf 77}, 063601 (2008).



\end{thebibliography}
\end{document}